\documentclass[twocolumn,showpacs,preprintnumbers]{revtex4}
\usepackage{amssymb}

\usepackage[dvips]{graphicx}

\begin{document}

\title{Ensembles of plasmonic nanospheres at optical frequencies and a problem of
negative index behavior}
\author{E.V. Ponizovskaya and A.M. Bratkovsky}

\affiliation{Hewlett-Packard Laboratories, 1501 Page Mill Road, Palo
Alto, California 94304}

\date{\today}

\begin{abstract}

Arrays of metallic nanoparticles support individual and collective
plasmonic excitations that contribute to unusual phenomena like
surface enhanced Raman scattering, anomalous transparency, negative
index, and subwavelength resolution in various metamaterials. We
have examined the electromagnetic response of dual Kron's lattice
and films containing up to three monolayers of metallic nanospheres.
It appears that open cubic Kron's lattice exhibits `soft'
electromagnetic response but no negative index behavior. The
close-packed arrays behave similarly: there are plasmon resonances
and very high transmission at certain wavelengths that are much
larger than the separation between the particles, and a `soft'
magnetic response, with small but positive effective index of
refraction. It would be interesting to check those predictions
experimentally.

\end{abstract}

\maketitle


\section{Introduction}

Media with strong dispersion of the refractive index may support backward
waves, as is obvious from the following relation established by Lord
Rayleigh in 1877:
\begin{equation}
n_{g}=n-\lambda \frac{dn}{d\lambda }
\end{equation}
where $n_{g}$ is the group index for waves with wavelength $\lambda ,$ phase
velocity is $v_{g}=c/n_{g}$, group velocity $v=c/n,$ where $c$ is the light
speed in vacuo. Hence, the group velocity may become negative in a system
with large positive dispersion. The structures that support backward waves
have been known since early 1900s and widely used in antenna and electronics
technologies since 1950s. Media where the directions of phase and group
velocities are opposite are known to produce negative refraction\cite{mand44}%
. It was noted in Ref.~\cite{mand44} that e.g. dispersion of light close to
excitonic frequencies in solids can be negative. It is also true of
artificial metamaterials with strong spatial dispersion, like photonic
crystals \cite{silin,kosaka98}. Pafomov \cite{pafomov59} and Veselago \cite
{ves67} showed that backward waves can propagate in isotropic medium with
simultaneously negative permittivity $\epsilon $ and permeability $\mu $,
exhibiting negative refraction, inverse Doppler and Vavilov-Cherenkov
effects. Currently, such metamaterials are called negative index
metamaterials (NIM). Veselago has noticed that a slab of NIM with thickness $%
d$ would work as a \emph{flat lens}. The lens could not produce an image of
a distant object, since it cannot focus a parallel beam of light, but
produces a replica of the object if it is placed less than distance $d$ away
from the nearest surface of the slab.

Recently, unusual properties of NIM systems attracted considerable attention
followed first theoretical \cite{pen96wire,pen99srr} and then experimental
\cite{smith01}\ demonstration of feasibility of negative refraction by
periodic metallic metamaterials like periodic wire meshes producing negative
$\epsilon $ \cite{pen96wire} and split-ring resonators (SRR) giving negative
$\mu $ \cite{pen99srr} in microwave region of incident radiation. Pendry has
showed theoretically that ideal Veselago lens can produce sub-wavelength
resolution (\emph{perfect lens}) \cite{pen00}. The effect appears because
ideal NIM slab supports surface plasmon modes that are in resonance with
incident radiation at any angles of incidence\cite{haldane02,gar02}. The
incident field pumps those plasmon modes up and this lead to enhancement of
evanescent waves reaching the surface on the image side of the slab further
from the source. The induced displacement currents re-emit the light that
reconstructs the image of the source without loss of resolution for features
smaller than $\lambda $. Similar effect involving surface plasmons is also
responsible for extraordinary transmission in thin metallic hole arrays\cite
{ebb04}. This behavior is quite fragile, however, and limited by losses and
spatial dispersion (metamaterials granularity)\cite
{haldane02,smithapl03,bnim05}. Nevertheless, it has indeed been demonstrated
experimentally by Lagarkov and Kissel that the NIM slab built with SRR
interspersed with the wire mesh is able to resolve features $\sim \lambda /6$
in the source separated by for microwave radiation ($f=1.7$GHz)\cite{lagar04}%
. Later on, the sub-wavelength resolution ($\sim \lambda /6)$ was
demonstrated by N. Fang \textit{et al}. in the visible range using
silver slab as a plasmonic medium\cite{nick05}. The silver slab is
not a material with both permittivity and permeability negative
(\emph{double negative)}, it only has $\epsilon <0$ as any metal at
frequencies below plasmon frequency, and $\mu >0,$ but the effect is
still possible because the system is in quasistatic limit where the
sign of permeability $\mu $ drops out of the result for image
intensity\cite{pen00}. Various papers describe metamaterials that
show NIM-like response at far infrared frequencies\cite {linden04}
and, most interestingly, in near-optical and optical interval\cite
{szha1,szha2,vlad05,dol06,wei06}.

\section{Electromagnetic response of metallic nanoparticle assemblies}

The metallic periodic systems like `fishnet' structure \cite{szha1} support
an infinite set (bands) of electromagnetic (EM) waves, $\omega =\omega
_{k}^{n}$, where $k$ is the wavevector (quasi-momentum), $n=1,2,\dots $ the
number of the band (Floquet mode). At small wavevectors $k$ such a crystal
can be characterized by effective permittivity $\epsilon $ and permeability $%
\mu $\cite{efros03}. Because of strong dispersion, it is easy to find
crystals where some of the higher bands (second or third) support backward
waves corresponding to negative group velocity, see Refs.\cite
{silin,kosaka98}. If such a band exists alone in a particular (usually
narrow) frequency range, the crystal would operate as a NIM at those
frequencies. Eleftheriades \textit{et al}. have considered recently a
possibility of an isotropic 3D NIM crystals (3D transmission lines) that
might support backward waves in the first band if made of lumped capacitors
and inductances, Ref.\cite{elef05}, see Fig.~1a. This is directly follows
from the 3D transmission line (TL) model suggested by Kron for Maxwell
equations in isotropic space with positive permittivity and permeability, $%
\epsilon ,\mu >0$\cite{kron43}. In the case of `positive' medium the
dispersion at small $k$-vectors is positive for the lowest energy
band, where eigenfrequency $\omega =\omega _{k}$ is a growing
function of $k$. It should become a decreasing function of $k$ after
a dual transformation of the TL. Indeed, this is the standard
passband-to-stopband transformation,
where the dispersion remains exactly the same after a substitution $%
C\rightleftharpoons L$ with obvious replacement $\omega
\rightleftharpoons \omega ^{\prime }=\omega ^{-1}$, which means that
the group velocity for the dual band changes sign: $v_{g}=d\omega
/dk>0\rightarrow v_{g}^{\prime }(\omega ^{\prime })<0.$ The dual
transformation does indeed result in negative dispersion in the
doubly degenerate first band around the $\Gamma$ point
\cite{elef05}. However, the Kron TL lattice contains a few elements
per unit cell and, as a consequence, there is also another
`spurious' band present in the first Brillouin zone. The 3D TL
crystal can be fairly well impedance matched to free space in
$\sim$GHz range and it supports a backward wave. The `spurious' band
can couple to highly attenuated evanescent waves, but losses may be
the limiting factor in subwavelength resolution experiment with dual
Kron's TL lattice rather than the coupling to the `spurious'
band\cite{elef05}.

It would be interesting to find an implementation of the dual Kron
3D transmission line model for NIM in visible range. With this in
mind, Engheta \emph{et al}. have speculated that Ag nanoparticles
can play a role similar to lumped inductance at optical frequencies.
One can try, for instance, substituting lumped inductances by Ag
nanoparticles in the dual Kron lattice, hoping that the capacitive
coupling between the particles would bring about the same band
structure as in 3D TL\cite{eng06}, see Fig.~1b. The nanoparticle
system in Fig.~1b is a poor representation of 3D TL system shown in
Fig.~1a: although one does have both electric and magnetic response
from metallic nanoparticles in ac field, they cannot be considered
as lumped circuit elements. Indeed, the interaction between
particles beyond nearest neighbors is likely important in open
structures like Kron's cubic lattice, in addition to their
non-negligible self-capacitance. We have calculated the two cases:
(i) lattice of Ag spheres in vacuo (dielectric constant of the
matrix is unity, $n_m=1$, and (ii) dielectric constant of the matrix
is $n_m=1.4$. The second case corresponds to experimentally
accessible case of metallic spheres embedded in an organic matrix.
We see from Fig.~1d that the realistic case of Ag spheres in the
dielectric matrix the electromagnetic response of the system is
`soft' (Re($n$) is small) but positive. Correspondingly, the dual
Kron's lattice of Ag nanoparticles in vacuo shows only a weak
negative index behavior, Re($n)\approx -0.3$ but it is overwhelmed
by losses, since Im$(n) \gg |$Re($n)|$. It is likely, therefore,
that the weak negative index behavior of Ag spheres in dual Kron's
lattice won't be observable in sub-wavelength focusing experiments.
\begin{figure}[tbp]
\includegraphics[width=0.5\textwidth]{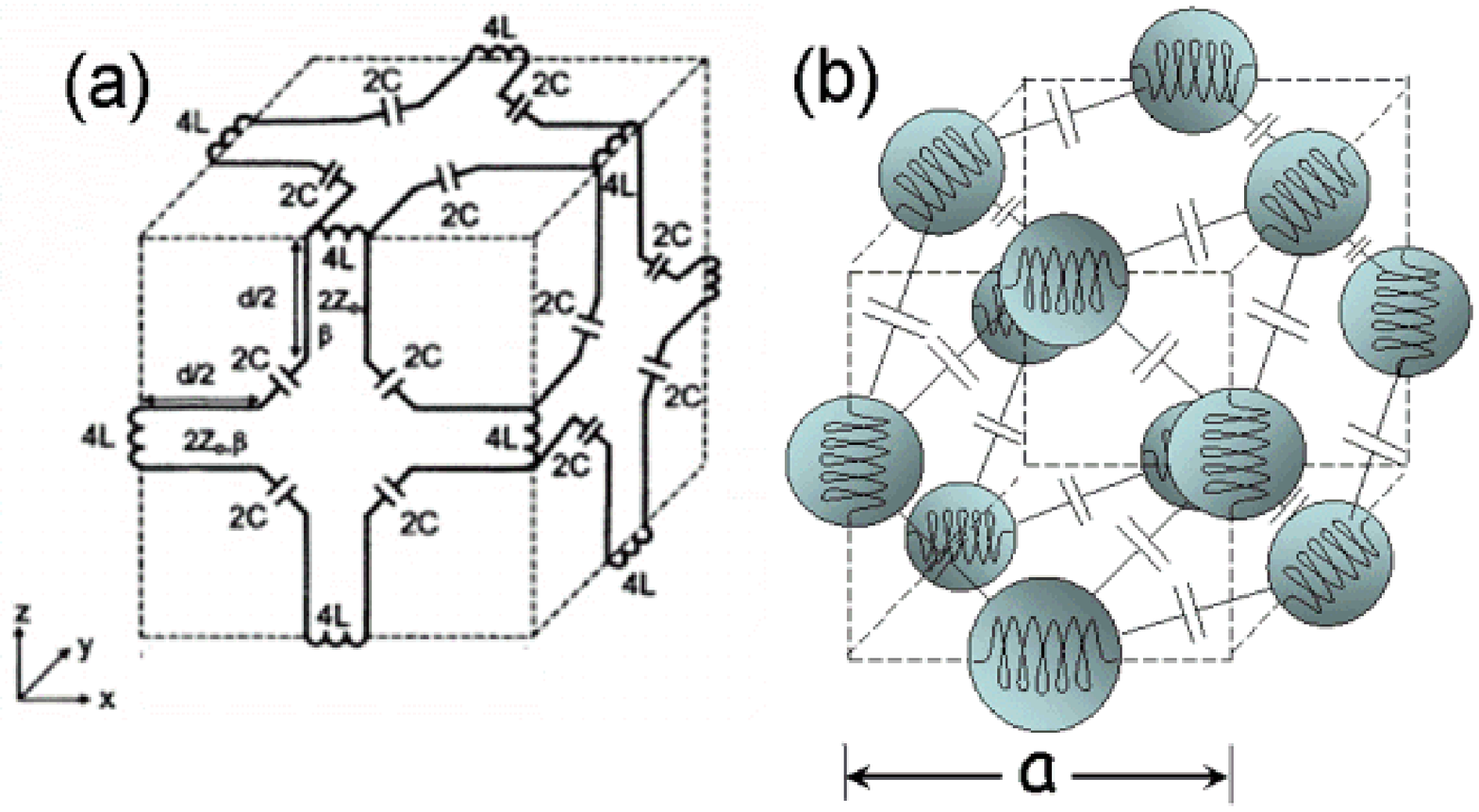} %
\includegraphics[width=0.4\textwidth]{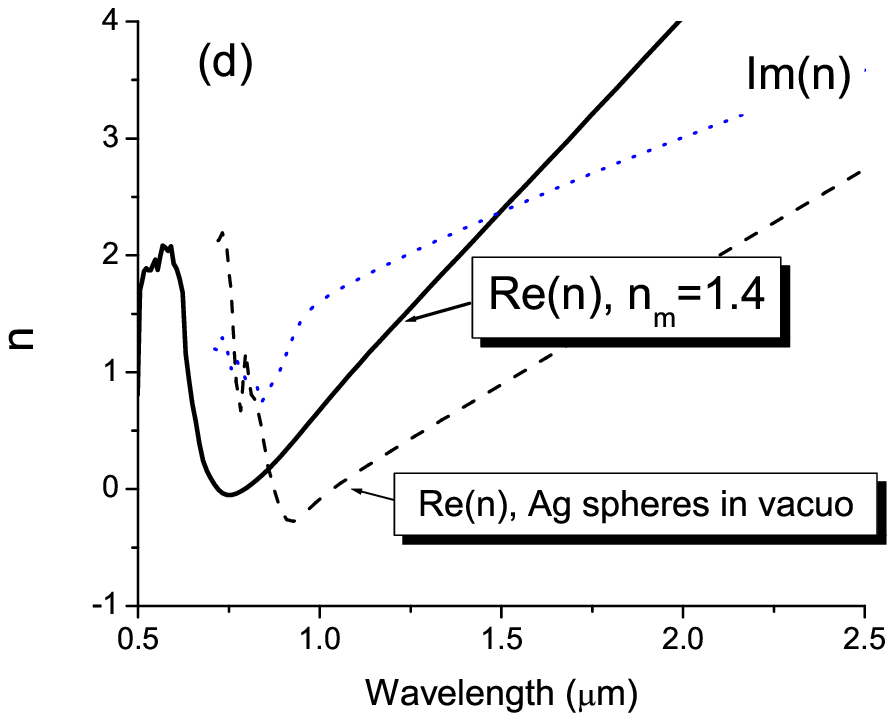} %
\includegraphics[width=0.4\textwidth]{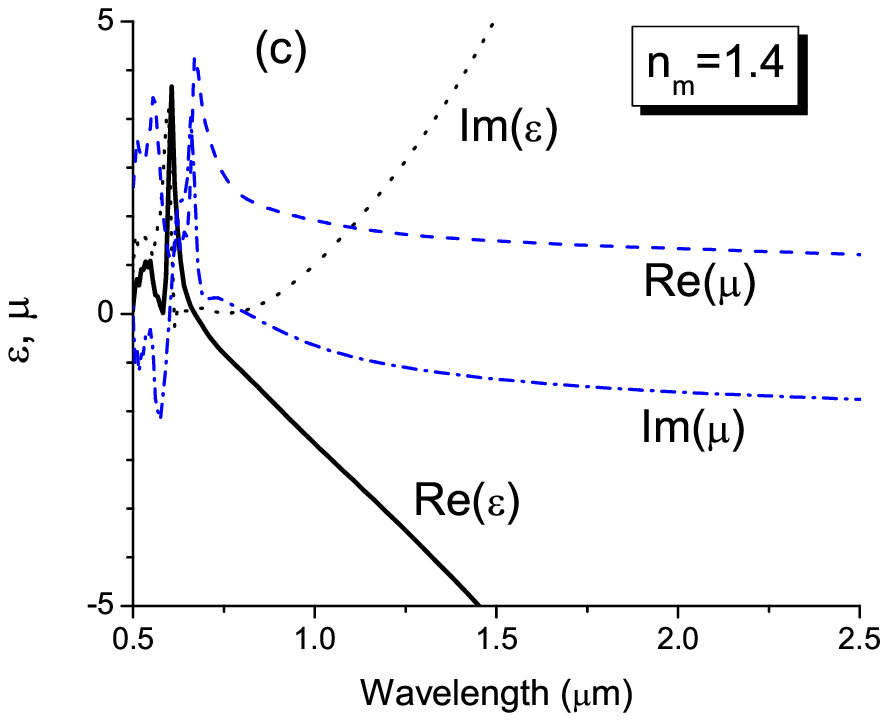}
\caption{ Schematic of dual Kron lattice \protect\cite{elef05} (a)
approximated by the cubic lattice of Ag nanoparticles with radius $R=100$~nm
and lattice period $a=300$nm (b). Particles themselves show some inductance
(marked by the coil sign) and are capacitively coupled to each other, as
schematically indicated in the graph (b). There is some resemblance to the
lattice of lumped elements\protect\cite{elef05} (a). (c) The effective
permittivity $\protect\epsilon (\protect\omega )$ and permeability $\protect%
\mu (\protect\omega )$ of a slab (2 u.c.) of Ag nanoparticles (b).
(d) Real and imaginary parts of index $n$, for Kron's lattice in
vacuo and in matrix material with $n_m=1.4$. Ag lattice in the
matrix shows some `softness' at $\lambda\approx 0.9 \mu$m [small
Re$(n)$], whereas system in vacuo exhibits  a weak negative index
behavior with Re($n)\approx -0.3$ but Im($n)\gg $ $|{\rm Re}(n)|$. }
\label{fig:kron}
\end{figure}

To get more insight into electromagnetic (EM) response of nanoparticle
arrays, we have performed extensive Finite Difference Time Domain (FDTD)
modeling\cite{fdtd} of dual Kron lattices (Fig.~1) and close packed layers
of spherical Ag nanoparticles (Figs.~2-4). The dielectric constant of Ag was
assumed to have a Drude form
\begin{equation}
\epsilon (\omega )=1-\frac{\omega _{p}^{2}}{\omega (\omega +i\gamma )},
\end{equation}
with $\omega _{p}=9.04$~eV, $\gamma =0.02$~eV \cite{johnson72}. The bare Mie
resonance of Ag nanoparticles would be at $\omega _{M}=\omega _{p}/\sqrt{3}%
=5.22$~eV or wavelength $\lambda _{M}=238$~nm\cite{stratton}. The
transmission characteristics of the slabs have been calculated by FDTD and
then used to estimate the effective index of refraction $n$, permittivity $%
\epsilon $ and permeability $\mu $ from complex scattering coefficients
according to a standard procedure\cite{defemu}.

The calculated effective permittivity and permeability of the dual Kron's
lattice of Ag particles with radius $R=100$nm show a series of sharp
resonances, Re($\epsilon )<0$ at $\lambda >0.8\mu $m, and Re($\mu )$ becomes
negative at $\lambda <1\mu $m. In the same region Re$(\epsilon)$ is also
negative, and we see that the material has a small negative index at $%
\lambda\approx0.9\mu$m, where Re$(n)\approx -0.3$. However, the losses are
very large there, Im$(n)\approx 1.7$, and this will likely preclude
sub-wavelength resolution with the lens made with this metamaterial. Large
losses seem to be quite general for systems of metallic nanoparticles\cite
{eng06} and may severely limit their use for the purpose of sub-wavelength
resolution.

It is easily understood that the nanospheres should be almost touching to
increase capacitive coupling between them. To see this, we considered Kron's
lattice with Ag nanoparticles with radius $R=300$nm with the gap of $300$nm
between then, embedded in a material with dielectric constant $\epsilon_m
=6.5$. In this case the index remained positive at all wavelengths of
interest, Re$(n)\approx 2.5$ at $\lambda=1.2-3.2\mu$m, since the
displacement current were ineffective in producing Re$(\mu)<0$, and Re$%
(\epsilon)$ only dipped below zero at a couple of very narrow intervals
around resonant wavelengths. Since the volume fraction of the metal was
quite small, the losses were minimal, typically Im$(n)<0.3$. To get NIM
behavior, one needs to sharply increase the metal fraction, but this
immediately leads to prohibitively large losses, as described above.

We have also studied arrays of Ag nanospheres with radius $R=30$nm close
packed into one, two, and three monolayers, Figs.~2-4. Such systems can be
prepared, for instance, by Langmuir-Blodgett technique. The particles have
been stacked in a triangular close packed monolayer (A), close packed
bilayer with AB packing (B is a triangular layer of nanospheres placed on
top of the A layer so that the top sphere touches three underlying spheres
in the layer A). Note that there are `see-through' channels of deep
sub-wavelength size in either A monolayer or AB bilayer that facilitate high
transmission of at $\lambda \sim 1.1\mu $m, Figs.~2-4. The effect is
apparently similar to extraordinary transmission of light through
sub-wavelength hole arrays in metallic films\cite{ebb04}. It is worth
comparing this with Kron's system, Fig.~1b, which is an open cubic lattice
and has high transparency even when a slab with 3 unit cells is considered.
The transmission remains high even for touching or slightly overlapping
nanoparticles, so the assembly is transparent and conductive at the same
time. Quite expectedly, the transmission of AB\ bilayer is reduced in
comparison with a monolayer, from $T\approx 80\%$ to about 60\%, but it
still remains very large.

\begin{figure}[tbp]
\includegraphics[width=0.4\textwidth]{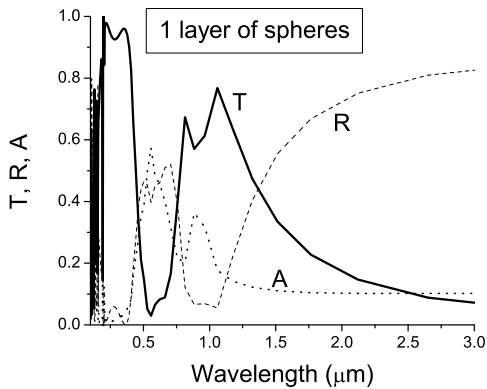} %
\includegraphics[width=0.4\textwidth]{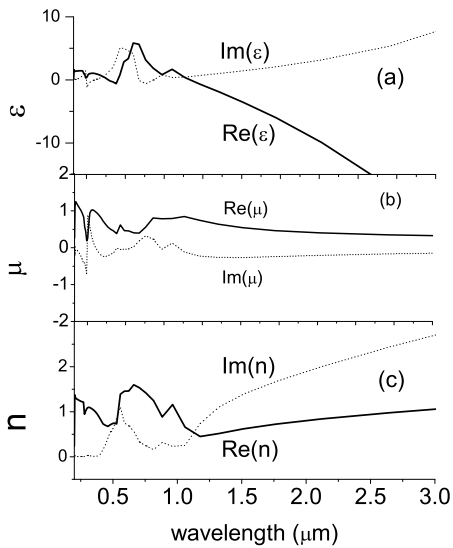}
\caption{(Top panel) Transmission $T$, reflection $R$, and absorption $A$ of
a monolayer of closely packed nanospheres with radius $a=30$nm. Note a very
large transmission $T\approx 0.75$ at $\protect\lambda $=1.1um, much larger
than the period of the monolayer, $\protect\lambda \gg a$. (Bottom panel)
Permittivity $\protect\epsilon $(a), permeability $\protect\mu $, and the
effective index $n$ of the monolayer of Ag nanoparticles. }
\label{fig:1L}
\end{figure}

System with three layers of nanospheres packed in ABC sequence
characteristic of face-centered cubic lattice (fcc) is interesting, because
there is no `see-through' channels in ABC stack. However, even in this case
transmission exceeds 40\%\ at the resonance, Fig.~4. We see from Figs.~2-4
that one to three monolayers of close packed metallic Ag nanospheres produce
an extraordinary transmission in the vicinity of $\lambda =1.1\mu $m, which
is much larger than the radius of the spheres $R$ and the lattice spacing $%
a. $ It is also much larger than the Mie resonant wavelength $\lambda
_{M}=238$~nm. Since in ABC trilayers there is no open ``channels'' for light
to squeeze through the film, the explanation of unusual `transparent metal'
behavior lies in the fact that the incident light strongly couples to an
array.%
\begin{figure}[tbp]
\includegraphics[width=0.4\textwidth]{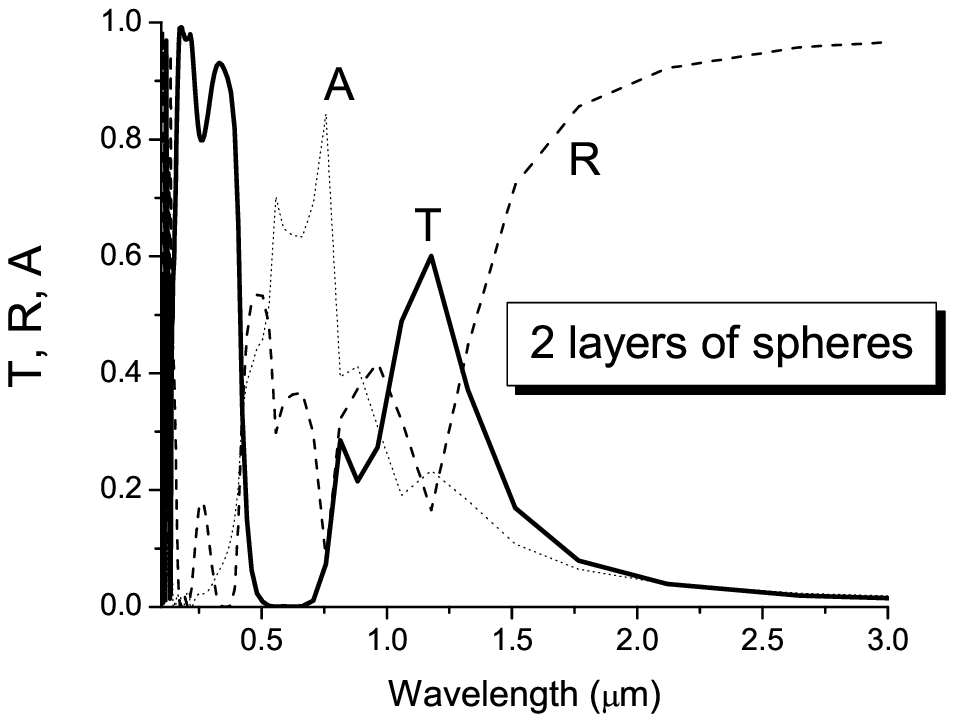} %
\includegraphics[width=0.4\textwidth]{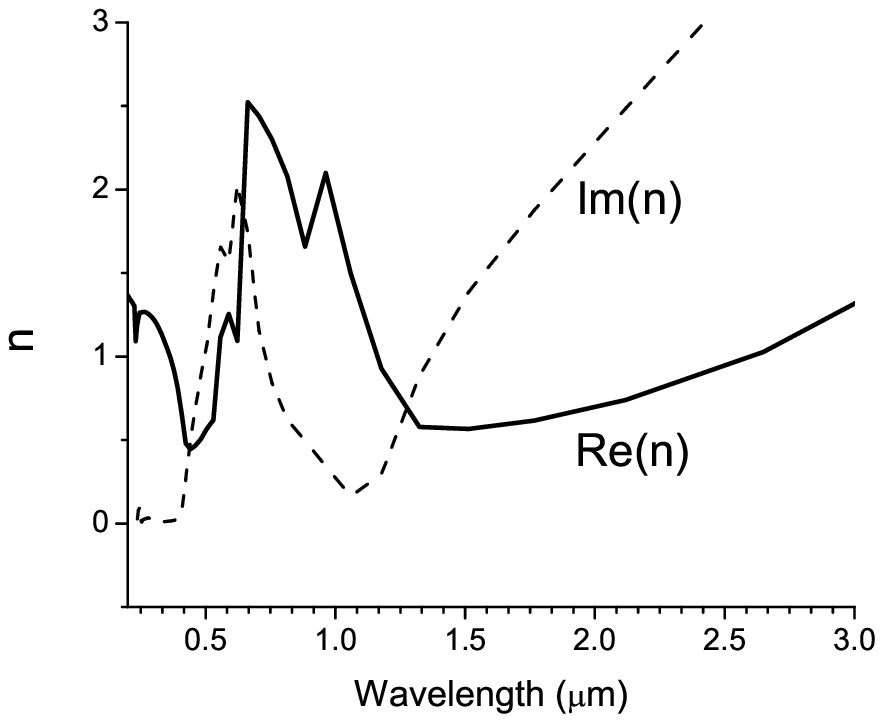}
\caption{(Top panel) Transmission $T$, reflection $R$, and
absorption $A$ of a two (AB packing sequence) layers of closely
packed nanospheres with radius $a=30$nm. Note a very large
transmission at $\protect\lambda $=1.1um, much larger than the
period of the monolayer, $\protect\lambda \gg a$. (Bottom panel) The
effective index $n$ of the bi-layer of Ag nanoparticles. We
observe a transmission in excess of 50\% at $\protect\lambda $=1.1$%
\protect\mu $m.}
\label{fig:2L}
\end{figure}
Strong coupling of the incident light is obviously facilitated by
periodic `roughness' of the arrays, since the (quasi)momentum
conservation is easier to meet. It would be interesting to see how
surface plasmon polaritons are supported by arrays of nanoparticles,
in other words, what kind of surface plasmon waveguiding is possible
with arrays of nanoparticles considered in the present
paper\cite{ebb03}.

Importantly, electric field concentrates in the region where the
spheres touch, especially when the centers of the spheres are
oriented along a polarization of incident electric field, which
looks analogous to corresponding electrostatic problem of two close
metallic spheres. In fact, the local electric field enhancement
exceeds a factor of $\eta \sim 30$, which would facilitate strong
Raman signal if the species were positioned on the particles near
the field peaks (Raman enhancement factor $\eta^4\sim 10^6$).

In terms of electromagnetic response, there is a clear topological
difference between a monolayer and a few-layer systems. Indeed, at
normal incidence the magnetic field in the incident wave cannot
excite any displacement currents that would for a close loops, since
the field does not `see' any such loops. In AB and ABC layers there
are three-membered loops in the fcc lattice that can produce such a
magnetic response. As follows from the data in Figs.~3,4, we do see
some `softness' in magnetic response of the AB and ABC arrays in
Re($\mu )$ at $\lambda \lesssim 0.8\mu $m, Fig.~2b. At the same
time, the index for the monolayer is positive and does not show any
particular `softness', cf. Fig.~2.

\begin{figure}[tbp]
\includegraphics[width=0.4\textwidth]{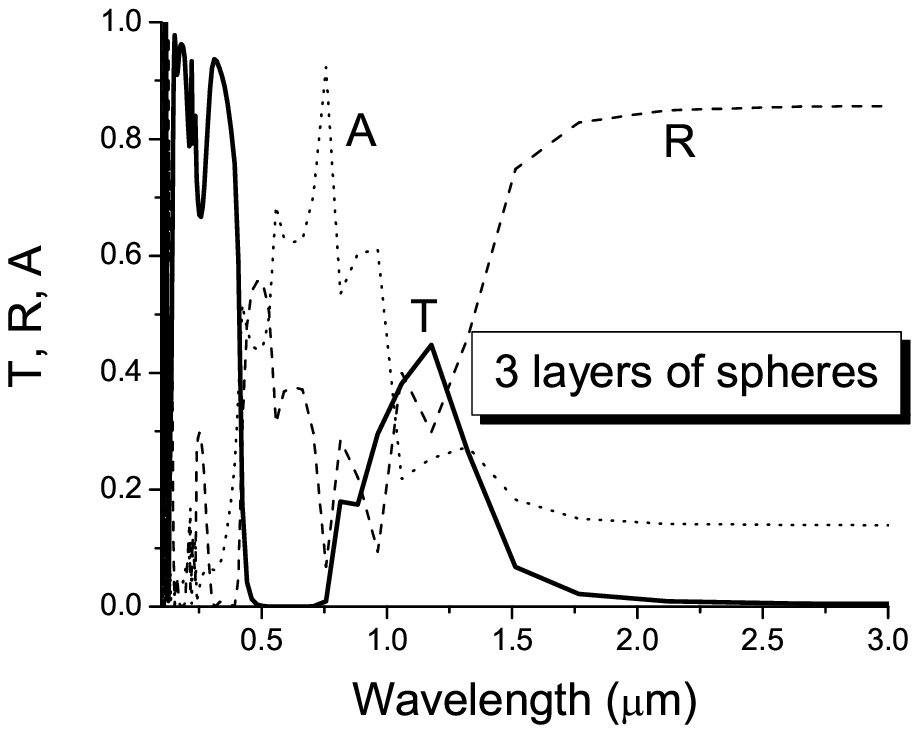} %
\includegraphics[width=0.4\textwidth]{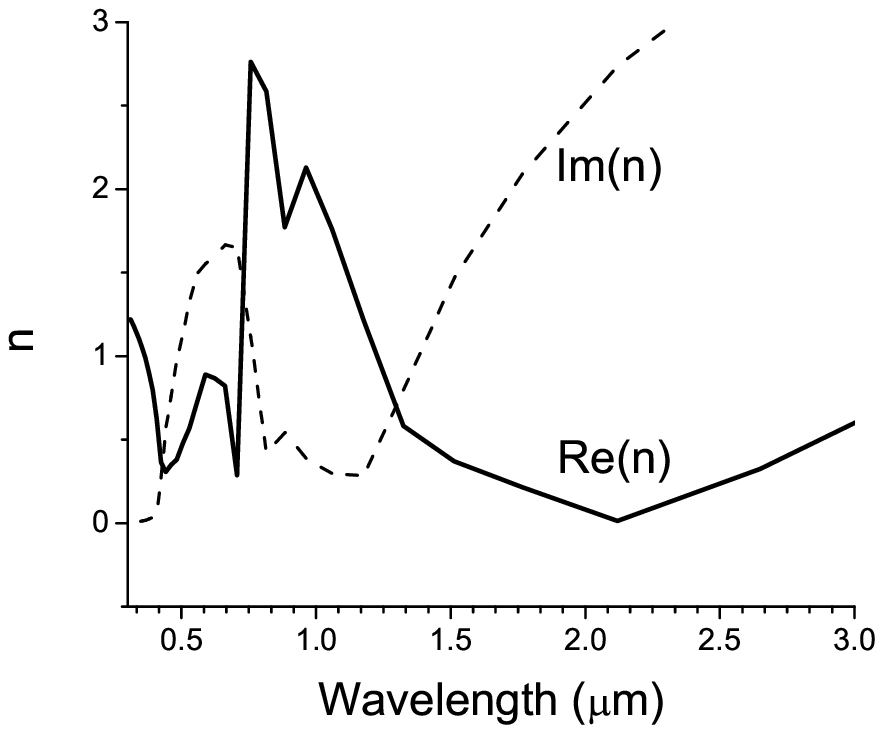}
\caption{(Top panel) Transmission $T$, reflection $R$, and
absorption $A$ of a three (ABC packing sequence) layers of closely
packed nanospheres with radius $a=30$nm. (Bottom panel) The
effective index $n$ of the bi-layer of Ag nanoparticles. Although
there is no `see-through' channels in this
structure, we still observe a large transmission at $%
\lambda $=1.1$\mu $m.} \label{fig:3L}
\end{figure}

\section{Conclusions}

We have analyzed various arrays of metallic nanoparticles in search for an
isotropic negative index behavior that was envisaged for dual Kron's cubic
lattice of lumped circuit elements (3D transmission line model)\cite{elef05}%
. In particular, we wanted to see if the assembly of metallic
nanoparticles can exhibit effective negative index
behavior\cite{eng06}. We have found that dual Kron's lattice of
closely spaced Ag nanoparticles does not exhibit negative index
behavior while embedded in a matrix, as we have shown for a matrix
with $n_m=1.4$. Kron's lattice in vacuo may show a weak negative
index behavior. However, it is overwhelmed by losses that would
preclude sub-wavelength resolution with a slab of this metamaterial.
We also made an extensive FDTD study of close-packed fcc-like arrays
of Ag nanospheres. Apparently, they do support collective plasmon
excitations and extraordinary transparency. Even in the opaque
fcc-like trilayer of Ag nanospheres, where there are no see-through
channels for light, the plasmons do transfer the excitation from the
front to back surface and re-emit it, providing for high
transparency in excess of 40-50\%. In a monolayer of Ag particles
the transparency is in excess of 80\%. Interestingly, the monolayer
does not show electromagnetic `softness' in response, whereas in
both bi- and tri-layers (AB and ABC packed films) we have found that
Re$(n)\approx 0.3$ at $\lambda \approx 0.7\mu$m. Although the arrays
do not behave as negative index media, we see an intriguing behavior
characteristic of a `transparent metal'. All those predictions would
be very interesting to test experimentally.


\begin{references}

\bibitem{mand44}  L.I. Mandelshtam, \emph{Lectures in Optics, Relativity, and Quantum Mechanics} (Moscow,
Nauka, 1972), p. 389.

\bibitem{silin} R.A. Silin, Uspekhi Fiz. Nauk \textbf{175}, 562 (2006);
R.A. Silin and V.P. Sazonov, Delay Systems (Radio, Moscow, 1966).

\bibitem{kosaka98} H. Kosaka, T. Kawashima, A. Tomita, M. Notomi, T. Tamamura, T.
Sato, and S. Kawakami, Phys. Rev. B \textbf{58}, 10096(R) (1998); M.
Notomi, Phys. Rev. B \textbf{62}, 10696 (2000).

\bibitem{pafomov59} V.E. Pafomov, Zh. Eksp. Teor. Fiz. \textbf{36}, 1853 (1959)
\bibitem{ves67} V.G. Veselago, Usp. Fiz. Nauk \textbf{92}, 517 (1967).
\bibitem{pen96wire} J.B. Pendry, A.J. Holden, W.J. Stewart, and I. Youngs, Phys. Rev. Lett. \textbf{76}, 4773 (1996).
\bibitem{pen99srr} J.B. Pendry, A.J. Holden, D.J. Robbins, and W.J. Stewart, IEEE Trans. Microwave Theory Tech.
\textbf{47}, 2075 (1999).
\bibitem{smith01}  R.A. Shelby, D.R. Smith, and S.Schultz, Science \textbf{292}, 77 (2001).

\bibitem{pen00}  J.B. Pendry, Phys. Rev. Lett. \textbf{85}, 3966 (2000).


\bibitem{haldane02}  F.D.M. Haldane, cond-mat/0206420.
\bibitem{gar02} N. Garcia and M. Nieto-Vesperinas, Phys. Rev. Lett. \textbf{88}, 207403 (2002).


\bibitem{ebb04} W.L. Barnes, W.A. Murray, J. Dintinger, E. Devaux, and T.W. Ebbesen
Phys. Rev. Lett. \textbf{92}, 107401 (2004);  H.J. Lezec, A.
Degiron, E. Devaux, R.A. Linke, L. Martin-Moreno, F.J. Garcia-Vidal,
T.W. Ebbesen, Science \textbf{297}, 820 (2002).



\bibitem{smithapl03} D.R. Smith, D. Schurig, M.Rosenbluth, S. Schultz, S. A. Ramakrishna, and J. B. Pendry, Appl.
Phys. Lett. \textbf{82}, 1506 (2003).
\bibitem{bnim05} A.M. Bratkovsky, A.Cano, and A.P.
Levanyuk, Appl. Phys. Lett. \textbf{87}, 103507 (2005).

\bibitem{lagar04} A. N. Lagarkov and V. N. Kissel, Phys. Rev. Lett. \textbf{92}, 077401
(2004).
\bibitem{nick05} N. Fang, H. Lee, C. Sun, and X. Zhang, Science \textbf{308},
534 (2005).
\bibitem{linden04} S. Linden \emph{et al.}, Science \textbf{306}, 1351 (2004).

\bibitem{szha1} S. Zhang \emph{et al}., Opt. Express \textbf{13}, 4922 (2005).
\bibitem{szha2} S. Zhang \emph{et al}., Phys. Rev. Lett. \textbf{95},
137404 (2005).
\bibitem{vlad05} V. M. Shalaev \emph{et al}., Opt. Lett. \textbf{30}, 3356 (2005);
V.P. Drachev \emph{et al}., Laser Phys. Lett. \textbf{3}, 49 (2006);
A. Grigorenko \emph{et al}., Nature \textbf{438}, 335 (2005).

\bibitem{dol06} G. Dolling \emph{et al}., Science \textbf{312}, 892 (2006).

\bibitem{wei06} W. Wu, E.Kim, E. Ponizovskaya \emph{et al}., cond-mat/0610352; Appl. Phys. A (to appear).

\bibitem{efros03} A. L. Efros and A. L. Pokrovsky, cond-mat/0308611
(2003).

\bibitem{elef05} A. Grbic and G.V. Eleftheriades, J. Appl. Phys. \textbf{98}, 043106
(2005).
\bibitem{kron43} G. Kron, Phys. Rev. \textbf{64}, 126 (1943).

\bibitem{eng06} A. Alu and N. Engheta, cond-mat/0609625.


\bibitem{fdtd} A. Taflove and S.C. Hagness, {\it Computational Electrodynamics } (Artech House, Boston, 2000).


\bibitem{johnson72} P.B. Johnson and R.W. Christy, Phys. Rev. B {\bf 6}, 4370 (1972).

\bibitem{stratton} J.A. Stratton, {\it Electromagnetic Theory}  (McGraw-Hill, New York, 1941).

\bibitem{defemu} D.R. Smith, S. Schultz, P. Markos, and C.M. Soukoulis, Phys. Rev. B \textbf{65}, 195104
(2002).

\bibitem{penram03} J.B. Pendry and S.A. Ramakrishna, Physica B \textbf{338}, 329 (2003).



\bibitem{ebb03} S.I. Bozhevolnyi, V.S. Volkov, E. Devaux, J.-Y. Laluet, T.W.
Ebbesen, Nature \textbf{440}, 508 (2006);
 W.L. Barnes, A. Dereaux, and T.W. Ebbesen, Nature \textbf{424}, 824 (2003);
T. Nikolajsen \emph{et al}., Appl. Phys. Lett. \textbf{85}, 5833
(2004).

\end{references}
\end{document}